# SEMI-SUPERVISED LEARNING-BASED SOUND EVENT DETECTION USING FREQUENCY DYNAMIC CONVOLUTION WITH LARGE KERNEL ATTENTION FOR DCASE CHALLENGE 2023 TASK 4

## Technical Report

*Ji Won Kim[1], Sang Won Son[1], Yoonah Song[1], Hong Kook Kim[1,2,*]*

[1] AI Graduate School, [2] School of EECS
Gwangju Institute of Science and Technology
Gwangju 61005, Korea
{{jiwon.kim, ssw970519, yyaass0531}@gm., hongkook@}gist.ac.kr

*Il Hoon Song[3], Jeong Eun Lim[3]*

[3]AI Lab., R&D Center
Hanwha Vision
Seongnam-si, Gyeonggi-do 13488, Korea
{ilhoon, je04.lim}@hanwha.com

## ABSTRACT

This report proposes a frequency dynamic convolution (FDY) with a large kernel attention (LKA)–convolutional recurrent neural network (CRNN) with a pre-trained bidirectional encoder representation from audio transformers (BEATs) embedding-based sound event detection (SED) model that employs a mean-teacher and pseudo-label approach to address the challenge of limited labeled data for DCASE 2023 Task 4. The proposed FDY with LKA integrates the FDY and LKA module to effectively capture time-frequency patterns, long-term dependencies, and high-level semantic information in audio signals. The proposed FDY with LKA–CRNN with a BEATs embedding network is initially trained on the entire DCASE 2023 Task 4 dataset using the mean-teacher approach, generating pseudo-labels for weakly labeled, unlabeled, and the AudioSet. Subsequently, the proposed SED model is retrained using the same pseudo-label approach. A subset of these models is selected for submission, demonstrating superior F1-scores and polyphonic SED score performance on the DCASE 2023 Challenge Task 4 validation dataset.

## 1. INTRODUCTION

The objective of sound event detection (SED) is to recognize and classify individual sound events originating from acoustic signals, along with their corresponding time stamps. The potential applications of the SED model extend beyond audio captioning [1] to various domains, such as wildlife tracking [2], equipment monitoring [3], and medical monitoring [4]. In recent years, SED has been extensively researched using deep learning models [5]. However, a significant challenge in using deep learning for SED is the requirement of strong labels, which are expensive and time-consuming. This problem has led to research on developing weakly supervised and semi-supervised learning techniques to overcome this challenge.

Last year, we proposed a selective kernel attention–residual convolutional recurrent neural network (CRNN) for the DCASE 2022 Challenge Task 4 and achieved fourth place in terms of PSDS Scenarios 1 (PSDS1) and 2 (PSDS2), with scores of 0.514 and 0.713, respectively, for the evaluation dataset. Nevertheless, the model developed last year encountered specific issues. Unlike images, two-dimensional (2D) audio data, like spectrograms, are not shift-invariant along the frequency axis. Employing conventional convolutional methods may potentially degrade the performance of SED, particularly for nonstationary events over time, such as a dog barking or a cat meowing. Additionally, relying only on averaging an attention mechanism can result in reduced performance due to the variability in event duration and repetitive sound events.

In this submission, we aim to improve the SED model proposed for the DCASE 2022 Challenge Task 4 by replacing the residual convolutional blocks with frequency dynamic convolution (FDY) [8] and large kernel attention (LKA) [9] blocks, which are referred to as FDY–LKA blocks. Compared to DCASE 2022 Task 4, DCASE 2023 Task 4 announced a new baseline using built-in bidirectional encoder representation from audio transformers (BEATs) embedding [7]. Consequently, this year's baseline performance was much better than last year's baseline due to better capturing high-level semantic information.

Motivated by this embedding approach using a pretrained model, we also employ BEATs embedding as the input feature for SED. In particular, the last FDY–LKA block outputs are concatenated with BEATs embedding and then they are inputted to a recurrent neural network (RNN). In addition, two-stage-based semi-supervised learning strategies are applied to train the proposed SED model with weakly labeled and unlabeled data. Per DCASE 2023 Task 4 rules, we submit an FDY–LKA-CRNN model without external datasets, three with BEATs embeddings, and four ensemble models from first and second stage based on the best PSDS1, PSDS2, sum of PSDS1 and 2 value, and all models.

[*] This work was supported in part by Hanwha Vision Co. Ltd., and by Institute of Information & communications Technology Planning & Evaluation(IITP) grant funded by the Korea government(MSIT) (No.2022-0-00963, Localization Technology Development on Spoken Language Synthesis and Translation of OTT Media Contents).

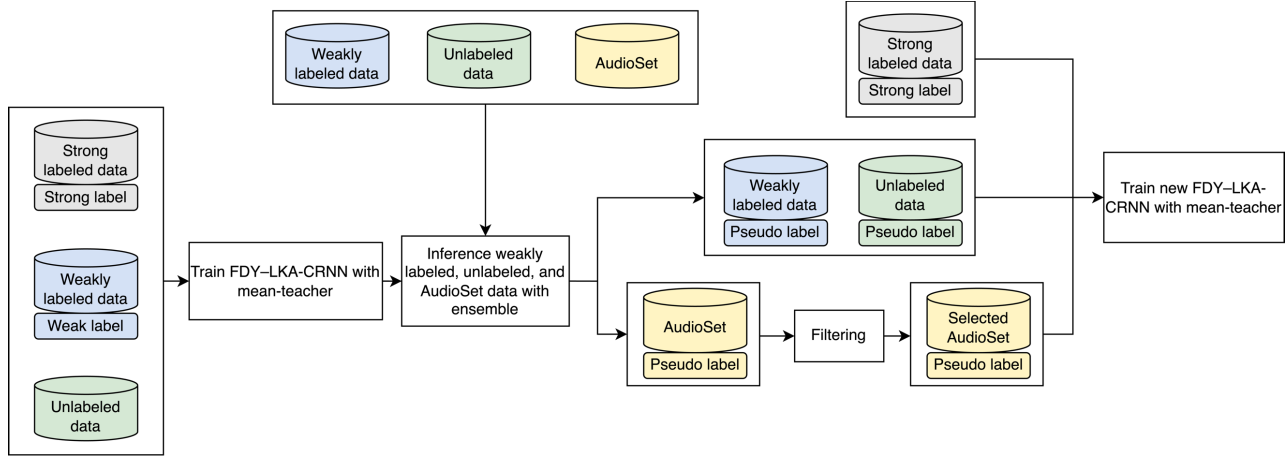

Figure 1: The training process for the proposed FDY–LKA-CRNN-based SED model, which entails a two-stage mean-teacher model comprising FDY–LKA-CRNN, trained through semi-supervised learning.

Following this introduction, Section 2 describes the dataset and input features of the SED model in this work. Section 3 proposes the FDY–LKA-CRNN with the BEATs embedding model and learning strategy. Then, Section 4 covers the experimental results of the suggested SED model on the DCASE 2023 Task 4 validation dataset. Finally, Section 5 concludes this report.

## 2. DATASET

The DCASE 2023 Challenge Task 4 comprises four datasets: weakly labeled data, unlabeled in-domain training data, strongly labeled synthetic data, and strongly labeled real data, with all audio clip data lasting 10 seconds. The strongly labeled synthetic dataset is distinct from the other datasets in that it is created by Scraper [11]. The weakly labeled dataset contains only class labels and is annotated for 1,578 clips. The unlabeled in-domain training dataset contains 14,412 audio clips. Finally, the real strongly labeled and synthetic datasets contain 3,470 and 10,000 clips, respectively. Besides the DCASE dataset, we employed AudioSet, which offers weakly labeled data. The DCASE and AudioSet datasets share overlapping labels, such as "blender," "cat," and "vacuum cleaner."

The following preprocessing procedures used the provided data as input to the model, starting by resampling the mono-channel signals from 44.1 to 16 kHz. Then, the audio signals are split into frames of 2,048 samples, each with a hop length of 160 samples. Each frame first performs a 2,048-point fast Fourier transform (FFT) employing 2,048 points, followed by a 128-dimensional mel-filterbank analysis. There are 1,001 frames for each 10 seconds audio clip. As a result, the input feature dimensions are 1001x128. The retrieved mel-spectrogram features are normalized using the mean and standard deviation for all training audio samples.

## 3. PROPOSED FDY–LKA-CRNN-BASED SED MODEL

Fig. 1 illustrates the training process of the proposed FDY–LKA-CRNN-based SED model, which employs a two-stage mean-teacher model. Further sections explain the proposed FDY–LKA-CRNN with BEAT embedding architecture and the two-stage semi-supervised learning.

### 3.1. Model architecture

Table 1 presents the proposed FDY–LKA-CRNN with BEAT embedding architecture, comprising one stem block, six FDY–LKA blocks, fusion block, and one RNN block. Initially, all input features of each audio clip are grouped to create a (1001×128×1)-dimensional spectral image as the input to the stem block. The shape of (x × y × z) indicates (frame × frequency × channel), and (x × y) means (frame × channel). The stem block consists of one convolutional block with 32 kernels for the first convolutional block. Convolution stem block has 3×3 kernels with a stride of 1×1 and received batch normalization, gated linear unit (GLU) activation, and a 2×2 average pooling layer.

Next, the output of the stem block is processed by the first FDY–LKA block that consists of a FDY, LKA module, batch normalization, GLU, and an average pooling layer, as presented in the table. Afterward, the output of each FDY–LKA block is passed to the next FDY–LKA block. Thus, the output of FDY–LKA-CNN becomes a (250×1×256)-dimensional feature map. The detailed network architecture of FDY–LKA blocks is described in the following section.

Along with FDY–LKA, we adopt embedding extracted by the BEATs encoder to use high-level semantic information properly. We employed average pooling or nearest neighbor interpolation to align the dimensions between the output of FDY–LKA-CNN and BEATs embedding. Both methods are applied in first and second stage for model diversity.

Then, this feature map is combined with the BEATs embedding, which has the dimensions (250×1×768). The combined feature map, with dimensions of (250×1×1024), is processed through a fully connected (FC) layer that reduces the channel size from 1024 to 256. The set of procedures executed to utilize BEATs embedding is referred to as Fusion Block. After then, this feature map is applied to the RNN block, consisting of two bidirectional gated recurrent units (Bi-GRUs) to learn the temporal context information, where a rectified linear unit (ReLU) is used as an activation function for each Bi-GRU.

Table 1. Network architecture of the proposed FDY–LKA-CRNN-based SED model, where the Fusion Block is performed when using BEATs embedding.

| Name | Layers | Output shape |
| --- | --- | --- |
| Input layer | Input : log-mel spectrogram | 1001×128×1 |
| Stem block | 3x3, Conv2D, @32 GLU, BN<br>2x2 average pooling layer | 500×64×32 |
| FDY–LKA blocks | FDY(K=4), @64, GLU, BN<br>LKA module<br>1x2 average pooling layer | 250×32×64 |
| | FDY(K=4), @128, GLU, BN<br>LKA module<br>1x2 average pooling layer | 250×16×128 |
| | FDY(K=4), @25, GLU, BN<br>LKA module<br>1x2 average pooling layer | 250×8×256 |
| | FDY(K=4), @256, GLU, BN<br>LKA module<br>1x2 average pooling layer | 250×4×256 |
| | FDY(K=4), @256, GLU, BN<br>LKA module<br>1x2 average pooling layer | 250×2×256 |
| | FDY(K=4), @256, GLU, BN<br>LKA module<br>1x2 average pooling layer | 250×1×256 |
| Fusion Block (optional) | Average pooling or interpolation on BEATs embedding | 250×768 |
| | Channel-wise Concatenation<br>(Output of FDY–LKA blocks (250, 256)<br>BEATs embedding (250, 768)) | 250×1024 |
| | Fully connected layer (1024, 256) | 250×256 |
| Recurrent neural network block | ( 256 Bi-GRU cells ) x 2 | 250×512 |

Finally, to attain a strong label for each audio clip, the (250×256)-dimensional output of the RNN block is processed by a FC layer and then a sigmoid function, resulting in a (250×10)-dimensional output, where 10 indicates the number of sound events to be detected. The attention layer is implemented to obtain a weak label by squeezing the frame dimensions.

### 3.2. FDY–LKA

The FDY applies a frequency-adaptive kernel to release the translation equivariance of the 2D convolution along the frequency axis for physical consistency with the time-frequency patterns in sound events. First, it extracts frequency-adaptive attention weights from input by applying average pooling over the time axis followed by two 1D convolutions along the channel axis. Between the two 1D convolutional layers, batch normalization and ReLU are applied. Finally, we apply an attention weight to the input feature map using an element-wise product operation.

The output feature map of FDY passes through the LKA module, effectively extracting attention maps and capturing long-term dependencies from relevant frames and channels. The LKA module consists of five steps: 1×1 convolution, Gaussian error linear unit (GELU), depth-wise convolution, depth-wise dilation convolution, and 1×1 convolution. Through the LKA module, the input feature map becomes the attention feature map, and self-attention is performed with an element-wise product operation between the input and output attention map. After the attention mechanism, the output feature map is batch-wise normalized and passes a convolutional block composed of 1x1 convolution, 3x3 depth-wise convolution, GELU, and a 1x1 convolutional layer.

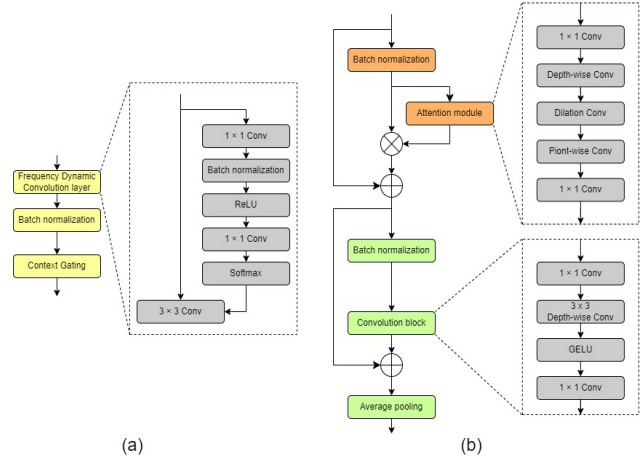

Figure 2: Structure of an FDY–LKA block composed of (a) a frequency dynamic convolution layer and (b) a large kernel attention module.

Fig. 2 illustrates the FDY–LKA block incorporating FDY and LKA module. As depicted, the average pooling layer is applied to the output of the feature map obtained using LKA.

### 3.3. Two-stage semi-supervised learning

The FDY–LKA-CRNN with the BEATs embedding-based SED model is trained using the mean-teacher approach based on a two-stage training procedure. After first stage, we generate strong pseudo-labels for the weakly labeled, unlabeled, and AudioSet datasets employed in second stage training.

We employ ensemble prediction derived from the student and teacher models to improve the quality of pseudo-labels, demonstrating the highest valid PSDS in the first, second, and combined scenarios. By thresholding the ensemble results by 0.5, we can obtain strongly labeled pseudo-labels for DCASE weakly labeled and unlabeled data. In the AudioSet, we apply applied the confidence score generated from weak predictions to select samples associated with DCASE classes. Through the confidence score and prediction results, strong pseudo-labels for the $C$-th class and $F$-th frame $pl_C^F$ are determined using the following equation:

$$pl_C^F = \begin{cases} 1, & \text{if } p_C^F > 0.5 \text{ and } p_C > 0.7 \text{ and } l_C = 1 \\ 0, & \text{otherwise,} \end{cases} \quad (1)$$

where $l_C$ denotes the weak label of the $C$-th class, $p_C$ represents the weak prediction value of the $C$-th class, and $p_C^F$ indicates the strong prediction value of the $C$-th class and $F$-th frame. This method allows strong pseudo-labels of weakly labeled AudioSet data for the next stage.

Table 2: Performance comparison of metrics of the baseline and versions of the proposed SED model on the validation dataset of the DCASE 2023 Challenge Task 4.

| Model | AudioSet | BEATs embedding | Ensemble | Event-based F1-score (%) | PSDS1 | PSDS2 |
|---|---|---|---|---|---|---|
| Baseline: CRNN-based mean-teacher model [13] | √ | Average pooling | – | 57.6 ± 0.7 | 0.4910 ± 0.003 | 0.7870 ± 0.007 |
| FDY–LKA-CRNN | – | – | – | 58.3 ± 1.3 | 0.4707 ± 0.029 | 0.7150 ± 0.039 |
| Stage 1 FDY–LKA-CRNN | – | Interpolation | – | 63.3 ± 0.3 | 0.5265 ± 0.004 | 0.7819 ± 0.002 |
| Stage 1 FDY–LKA-CRNN | – | Average pooling | – | 62.9 ± 1.0 | 0.5249 ± 0.004 | 0.7755 ± 0.009 |
| Stage 2 FDY–LKA-CRNN | √ | Interpolation | – | 63.4 ± 1.7 | 0.5427 ± 0.006 | 0.8056 ± 0.002 |
| Stage 2 FDY–LKA-CRNN | √ | Average pooling | – | 63.8 ± 0.9 | 0.5459 ± 0.004 | 0.8075 ± 0.002 |
| Stage 1&2 FDY–LKA-CRNN | √ | Both | √ | 65.6 ± 0.5 | 0.5667 ± 0.006 | 0.8154 ± 0.004 |

## 4. EXPERIMENTAL RESULTS

### 4.1. Model training

The FDY–LKA-CRNN-based SED model parameters were initialized in the first training stage using the Xavier initialization. The minibatch-wise adaptive moment estimation optimization technique was employed, involving decoupling the weight decay from the gradient-based updates. Additionally, a dropout method was applied to the FDY–LKA-CRNN model at a rate of 0.5. The learning rate was set according to the ramp-up strategy, with the maximum learning rate reaching 0.001 after 50 epochs. Several augmentation techniques, such as the time-frequency shift, time mask, mix-up, and filter augmentation, were applied to the training data. In the second stage, all training strategies remained identical to those used in first stage.

### 4.2. Discussion

The performance of the proposed SED model was evaluated using the measures defined in the DCASE 2023 Challenge Task 4 [12], such as an event-based F1-score and PSDS [6]. Table 2 compares the performance between the baseline with BEATs embedding and various versions of the proposed SED models on the validation dataset of the DCASE 2023 Challenge Task 4. As presented in the table, FDY–LKA-CRNN-based mean teacher model scored 0.7% higher, but 0.0203 and 0.072 lower in the F1-score, PSDS1, and PSDS2 than baseline with BEATS embedding achived. In comparison to the existing baseline, both FDY–LKA-CRNN interpolation and FDY–LKA-CRNN average pooling model have demonstrated superior performance, with increases in F1-score by 5.7% and 5.3% respectively, improvements of 0.0355 and 0.339 points in PSDS1, but 0.0051 and 0.0115 lower in the PSDS-scenario 2. Furthermore, the second stage interpolate model (single model) scored 0.1%, 0.0162, and 0.0237 higher on the F1-score, and the second stage averag PSDS1, and PSDS2, respectively, than the Stage 1 FDY–LKA-CRNN interpolation model. Also, the second stage average pooling model (single model) scored 0.9%, 0.021, and 0.032 higher on the F1-score, and the second stage averag PSDS1, and PSDS2, respectively, than the Stage 1 FDY–LKA-CRNN average pooling model.

Finally, we constructed ensemble models according to different model combinations between stages 1 and 2. Among various ensemble models, it was shown that an ensemble model composed of the Top 1–48 models improved F1-score, PSDS1, and PSDS2 by 8.0%, 0.0757, and 0.0284, respectively, compared to the baseline provided by DCASE 2023 Task 4.

## 5. CONCLUSION

This report presents an FDY–LKA-CRNN with a BEATs embedding-based SED model for DCASE 2023 Challenge Task 4, employing a pseudo-label and mean-teacher approach. The proposed FDY–LKA-CRNN with a BEATs embedding-based SED model was trained on the entire DCASE dataset in the first stage. Afterward, the trained proposed SED model generates pseudo-labeled data for unlabeled and weakly labeled data, including the AudioSet. This approach aimed to overcome the challenges posed by the lack of strong labels and to leverage the strengths of deep learning models for improved SED performance. By combining the FDY and LKA modules, we sought to effectively capture time-frequency patterns and long-term dependencies in the audio data. The proposed FDY-LKA-CRNN-based SED model was evaluated on the DCASE 2023 Task 4 validation dataset. The performance of various ensemble models derived from model checkpoints was also explored. The results demonstrated that an ensemble model comprising the Top 1–48 models in the F1-score achieved a 8.0% improvement, a 0.0757 increase in PSDS1, and a 0.0284 enhancement in PSDS2, compared to the baseline provided by DCASE 2023 Task 4.